\documentclass{PoS}

\usepackage{graphicx}                              % Used for includegraphics
\usepackage{amsmath,amsfonts}
\graphicspath{{./pics/}}                       % location for color fig.

\usepackage{epsfig}
\usepackage{amssymb}
\usepackage{amsfonts}
\usepackage{float}
\usepackage{bbm}
\usepackage{psfrag}

\newcommand{\preprintline}{\newline
\vskip -4.2cm
\rightline{\parbox{4cm}{\large\rm  HU-EP-07/52\\ DESY 07-173}}
\vspace{3.2cm}}

\title{The phase structure of a chirally invariant lattice Higgs-Yukawa model\preprintline}

\ShortTitle{The phase structure of a chirally invariant lattice Higgs-Yukawa model}

\author{\speaker{Philipp Gerhold}\\
        Humboldt-Universit\"at Berlin\\
        E-mail: \email{gerhold@physik.hu-berlin.de}}

\author{Karl Jansen\\
        DESY, Zeuthen\\
        E-mail: \email{Karl.Jansen@desy.de}}

\newcommand{\vs}{\vspace}
\newcommand{\hs}{\hspace}

\newcommand{\bdm}{\begin{displaymath}}
\newcommand{\edm}{\end{displaymath}}
\newcommand{\beq}{\begin{equation}}
\newcommand{\eeq}{\end{equation}}
\newcommand{\bea}{\begin{eqnarray}}
\newcommand{\eea}{\end{eqnarray}}
\newcommand{\bit}{\begin{itemize}}
\newcommand{\eit}{\end{itemize}}
\newcommand{\bc}{\begin{center}}
\newcommand{\ec}{\end{center}}
\newcommand{\re}{\relax{\rm I\kern-.18em R}}
\newcommand{\ID}{\mathbbm{1}}

\newcommand{\ie}{{\it i.e. }}
\newcommand{\D}{{\cal D}^{(ov)}}

\newcommand{\ImpSpace}{{\cal P}}

\newcommand{\sumFL}{\sum\limits_{i=1}^{N_f}}

\newcommand{\fermiMat}{{\cal M}}
\newcommand{\kapCrit}{\kappa_\mathrm{crit}}

\abstract{We consider a chirally invariant lattice Higgs-Yukawa model based on the Neuberger 
overlap operator $\D$. As a first step towards the eventual determination of Higgs 
mass bounds we present the phase structure of the model analytically 
in the large $N_f$-limit in the physically interesting region of the Yukawa coupling
constant.
We confront the analytically obtained phase diagram  with corresponding HMC-simulations
and find an excellent agreement at large values of $N_f$. In the opposite case the 
large $N_f$ computation still gives a good qualitative description of the phase diagram. 
We also present first and very preliminary results on the Higgs upper bound at one selected
cut-off $\Lambda$.}

\FullConference{The XXV International Symposium on Lattice Field Theory\\
                 July 30 - August 4 2007\\
                 Regensburg, Germany}

\begin{document}

\section{Introduction}
%---------------------
\label{sec:Introduction}

The main target of lattice studies of the Higgs-Yukawa sector of the 
electroweak standard model is the non-perturbative determination of the $\Lambda$-dependence
of the upper and lower bounds of the Higgs boson mass~\cite{Holland:2003jr,Holland:2004sd} 
as well as its decay properties, where $\Lambda$ denotes the cut-off of the theory. 
There are two main developments which 
warrant to reconsider these questions: first, with the advent of the LHC, 
we are to expect that properties of the standard model Higgs boson, such as 
the mass and the decay width, will be revealed experimentally. Second, there 
is, in contrast to the situation of earlier investigations of lattice 
Higgs-Yukawa  models~\cite{Smit:1989tz,Shigemitsu:1991tc,Golterman:1990nx,book:Jersak}, a consistent formulation of an Higgs-Yukawa model with an exact 
lattice chiral symmetry~\cite{Luscher:1998pq} based on the Ginsparg-Wilson 
relation~\cite{Ginsparg:1981bj}, which allows to establish a lattice version of chiral
symmetry while lifting the unwanted fermion doublers at the same time.

Before addressing the questions of the Higgs mass bounds and decay properties, we started
with an investigation of the phase structure of the model in order to obtain first
information about the region of the (bare) couplings in parameter space where eventual 
simulations of phenomenological interest should be performed.

In the present paper we basically summarize some of the most important results of our work on
the model's phase structure, which we have studied analytically in the large $N_f$-limit
for small as well as for large values of the Yukawa coupling constant~\cite{Gerhold:2007yb}, 
and numerically by means of HMC-simulations~\cite{Gerhold:2007gx}. 
%In the following section~\ref{sec:model} we
%introduce the investigated Higgs-Yukawa model and describe the applied HMC-algorithm. 
%The analytical results on the phase structure
%at small Yukawa coupling constants are then presented and compared to corresponding numerical
%results in section~\ref{sec:SmallYukawaCouplings}. 
Finally, we give a brief outlook towards some first and very preliminary results
on the upper Higgs boson mass obtained at one selected cut-off $\Lambda$.

\section{The model and its numerical treatment}
\label{sec:model}

The model, we consider here, is a four-dimensional, chirally invariant 
$SU(2)_L \times SU(2)_R$ Higgs-Yukawa model discretized on a finite lattice with $L$ 
lattice sites per dimension.
The model contains one four-component, real Higgs field $\Phi$ and $N_f$ fermion 
doublets represented by eight-component spinors $\psi^{(i)}$, $\bar\psi^{(i)}$, $i=1,...,N_f$
with the total action being decomposed into the Higgs action 
$S_\Phi$, and the fermion action $S_F$. It should be stressed here 
that {\it no gauge fields} are included within this model. 

The fermion action $S_F$ is based on the Neuberger overlap operator $\D$~\cite{Neuberger:1998wv}
and can be written as
\beq
S_F = \sumFL\bar\psi^{(i)} \Bigg[ \underbrace{\D + y_N B \cdot \left(\ID - \frac{1}{2\rho}\D  \right)}_{\fermiMat}\Bigg]
\psi^{(i)},
\quad
B_{n,m} = \ID_{n,m}\left[\frac{(1-\gamma_5)}{2}\phi_n 
+ \frac{(1+\gamma_5)}{2}\phi^{\dagger}_n \right].
\label{eq:DefYukawaCouplingTerm}
\eeq
It describes the propagation of the fermion fields as well as their coupling
to the Higgs field $\Phi$ through the Yukawa coupling matrix $B_{n,m}$ and the
Yukawa coupling constant $y_N$.
Here the Higgs field $\Phi_n$ was rewritten as a 
quaternionic, $2 \times 2$ matrix 
$\phi_n = \Phi_n^0\ID -i\Phi_n^j\tau_j$, with $\vec\tau$
denoting the vector of Pauli matrices,
acting on the $SU(2)$ index of the fermionic doublets.

Note that in absence of gauge fields the Neuberger Dirac operator can be
trivially constructed in momentum space, since for that
case its eigenvalues $\nu^\epsilon(p)$, $\epsilon=\pm 1$ for the allowed 
four-component momenta $p \in \ImpSpace$ are explicitly known. This will 
be exploited in the numerical construction of the overlap operator.

The model then obeys an exact, but lattice modified, chiral 
symmetry according to
\beq
\delta\psi^{(i)}=i\epsilon\gamma_5\left(1-\frac{1}{\rho} \D  \right)\psi^{(i)},\quad 
\delta\phi =2i\epsilon\phi,\quad
\delta\bar\psi^{(i)}=i\epsilon\bar\psi^{(i)}\gamma_5 ,\quad 
\delta\phi^\dagger =-2i\epsilon\phi^\dagger 
\label{eq:LatChiralSymmPhysField}
\eeq
recovering the chiral symmetry in the continuum 
limit~\cite{Luscher:1998pq}.

The lattice Higgs action $S_{\Phi}$ is given by the usual lattice notation
\beq
\label{eq:LatticePhiAction}
S_\Phi = -\kappa_N\sum_{n,\mu} \Phi_n^{\dagger} \left[\Phi_{n+\hat\mu} + \Phi_{n-\hat\mu}\right]
+ \sum_{n} \Phi^{\dagger}_n\Phi_n + \lambda_N \sum_{n} \left(\Phi^{\dagger}_n\Phi_n - N_f \right)^2
\eeq
with the only particularity that the fermion generation number $N_f$ appears in the quartic
coupling term which is a convenient convention for the large $N_f$ analysis. However, this
version of the lattice Higgs action is equivalent to the usual continuum notation~\cite{Gerhold:2007gx}.

For the numerical treatment of the model  
we have implemented an Hybrid-Monte-Carlo (HMC) algorithm
for {\it even} values of $N_f$
with $N_f/2$ complex pseudo-fermionic fields $\omega_j$ according to the HMC-Hamiltonian
\beq
H(\Phi,\xi,\omega_j) = S_\Phi[\Phi] + \frac{1}{2}\xi^\dagger \xi 
+ \sum\limits_{j=1}^{N_f/2}\frac{1}{2}\omega_j^\dagger \left[\fermiMat\fermiMat^\dagger\right]^{-1}\omega_j 
\eeq
where $\xi$ denotes the real momenta, conjugate to the Higgs field $\Phi$.
Since we focus here on checking the validity of our analytical investigation of the 
phase structure, which was determined in the large $N_f$-limit, the restriction to 
even $N_f$ does no harm. For the further details of this HMC algorithm we refer the
interested reader to Ref.~\cite{Gerhold:2007gx}.

The observables we will be using for exploring the phase structure 
are the {\it magnetization} $m$ and the {\it staggered magnetization} $s$, 
\begin{equation} 
m = \left[\sum\limits_{i=0}^3\Big|\frac{1}{L^4}\sum\limits_{n} \Phi_n^i\Big|^2\right]^{\frac{1}{2}}
\label{magnetizations} , \quad
s = \left[\sum\limits_{i=0}^3\Big|\frac{1}{L^4}\sum\limits_{n} \left(-1\right)^{\sum\limits_\mu n_\mu}\cdot\Phi_n^i\Big|^2\right]^{\frac{1}{2}}
\end{equation}
and the corresponding susceptibilities $\chi_m =  L^4\cdot \left[\langle m^2\rangle -\langle m \rangle^2 \right]$ and $\chi_s =  L^4\cdot \left[\langle s^2\rangle -\langle s \rangle^2 \right],$
where $\langle ... \rangle$ denotes the average over the $\Phi$-field 
configurations generated in the
Monte-Carlo process.

To locate the phase transition points,
we decided to fit the data for the susceptibilities  $\chi_m$, $\chi_s$
as a function of $\kappa_N$ 
according to the -- partly phenomenologically motivated -- ansatz
\begin{equation} 
\label{finitesizesus}
\chi_{m,s} = A_1^{m,s}\cdot\left(\frac{1}{L^{-2/\nu} + A_{2,3}^{m,s}(\kappa_N-\kapCrit^{m,s})^2}\right)^{\gamma/2},
\end{equation} 
where $A_1^{m,s}$, $A_{2,3}^{m,s}$, and $\kapCrit^{m,s}$ are the fitting parameters 
for the magnetic susceptibility and staggered susceptibility, respectively, 
and $\nu$, $\gamma$ denote the critical exponents of the $\Phi^4$-theory. 
Here $A_{2,3}^m$ ($A_{2,3}^s$) is actually meant to refer to
two parameters, namely $A_2^m$ ($A_2^s$) for $\kappa_N<\kapCrit^m$ ($\kappa_N<\kapCrit^s$) 
and $A_3^m$ ($A_3^s$) in the other case, such that 
the resulting curve is not necessarily symmetric. The phase transition point is then given at the 
value of $\kappa_N=\kapCrit^m$ ($\kappa_N=\kapCrit^s$) where the magnetic (staggered) susceptibility 
develops its maximum.

\section{Large $N_f$-limit for small Yukawa coupling parameters}
\label{sec:SmallYukawaCouplings}
The phase structure of the considered Higgs-Yukawa model can be 
accessed in the large $N_f$-limit by scaling the coupling constants 
and the Higgs field itself according to 
\beq
\label{eq:LargeNBehaviourOfCouplings1}
y_N = \frac{\tilde y_N}{\sqrt{N_f}}\,,\quad  
\lambda_N = \frac{\tilde \lambda_N}{N_f}\,,\quad
\kappa_N = \tilde \kappa_N\,,\quad
\Phi_n = \sqrt{N_f} \cdot\tilde\Phi_n\,,
\eeq
where the quantities $\tilde y_N$, $\tilde \lambda_N$, $\tilde \kappa_N$,
and $\tilde\Phi_n$ are kept constant in the limit $N_f\rightarrow\infty$
allowing to factorize the fermion generation number $N_f$ out of the 
effective action $S_{eff}[\Phi] = S_\Phi -N_f\log\det(\fermiMat)$.

One is thus left with the problem of finding the absolute minima of 
$S_{eff}[\Phi]$ 
in terms of the latter quantities. For sufficiently small values of 
the Yukawa and quartic coupling constants 
the kinetic term of the Higgs action becomes dominant allowing to restrict 
the search for the absolute minima of $S_{eff}[\Phi]$ to the ansatz
\beq
\label{eq:staggeredAnsatz}
\Phi_n = \hat\Phi \cdot \sqrt{N_f} \cdot \left(\tilde m + \tilde s\cdot (-1)^{\sum\limits_{\mu}n_\mu}    \right), \quad \hat\Phi\in \re^4, \quad |\hat\Phi|=1
\eeq
taking only a magnetization $\tilde m$ and a staggered magnetization $\tilde s$ into account.
After some work, which was presented in detail in \cite{Gerhold:2007yb}, one finally finds
for the effective action
\bea
S_{eff}[\Phi]
&=&  - N_f\cdot\sum\limits_{p\in\ImpSpace}
\log\Bigg[\left(\left|\nu^+(p)\right|\cdot \left|\nu^+(\wp)\right| 
+\frac{\tilde y_N^2}{4\rho^2} \left(\tilde m^2 - \tilde s^2\right) \cdot 
\left|\nu^+(p)-2\rho\right| \cdot \left|\nu^+(\wp)-2\rho \right|\right)^2\nonumber \\
&+&\tilde m^2\frac{\tilde y_N^2}{4\rho^2}  
\Big(\left|\nu^+(p)-2\rho\right|\cdot \left|\nu^+(\wp) \right| 
- \left|\nu^+(\wp)-2\rho\right| \cdot \left|\nu^+(p) \right|\Big)^2 
\Bigg]^2 + S_\Phi[\Phi],
\label{eq:EffActionRewrittenForSmallYFinal}
\eea
while the Higgs action in this setting reads
\bea
S_{\Phi} 
&=& N_f \cdot L^4\cdot \Bigg\{-8\tilde\kappa_N \Big(\tilde m^2-\tilde s^2\Big) +  \tilde m^2+\tilde s^2 
+\tilde\lambda_N \Big(\tilde m^4 +\tilde s^4 + 6\tilde m^2\tilde s^2 -2\left(\tilde m^2+\tilde s^2 \right) \Big)\Bigg\}.
\eea

The resulting phase structure in the large $N_f$-limit can then be determined by 
minimizing the effective action with respect to $\tilde m$ and $\tilde s$.
It is presented in Fig.~\ref{fig:PhaseDiagrams1}a for the selected value
of the quartic coupling constant $\tilde\lambda_N=0.1$ and $L=\infty$.
Here we distinguish between the following four phases:\\
\begin{tabular}{rlcrl}
{\it (I)} &  Symmetric (SYM): $\tilde m=0,\, \tilde s=0$
&&{\it (II)} & Ferromagnetic (FM): $\tilde m\neq 0,\, \tilde s=0$   \\
{\it (III)} & Anti-ferromagnetic (AFM): $\tilde m=0,\, \tilde s\neq 0$  
&&{\it (IV)} & Ferrimagnetic (FI): $\tilde m\neq 0,\, \tilde s\neq 0$   \\
\end{tabular}\\

In Fig.~\ref{fig:PhaseDiagrams1}b we compare this analytically obtained $N_f=\infty$,
$L=\infty$ phase structure with the results of corresponding HMC-simulations performed
on $8^4$- and $6^4$-lattices at $N_f=10$.
As expected we observe a good qualitative agreement between the numerical and analytical
results. On a quantitative level, however, the encountered deviations in Fig.~{\ref{fig:PhaseDiagrams1}}b 
need to be further addressed. These deviations
can be ascribed to finite volume effects as well as finite $N_f$ corrections.

\begin{figure}[htb]
\begin{tabular}{cc}
\hs{15mm}
\includegraphics[width=0.32\textwidth]{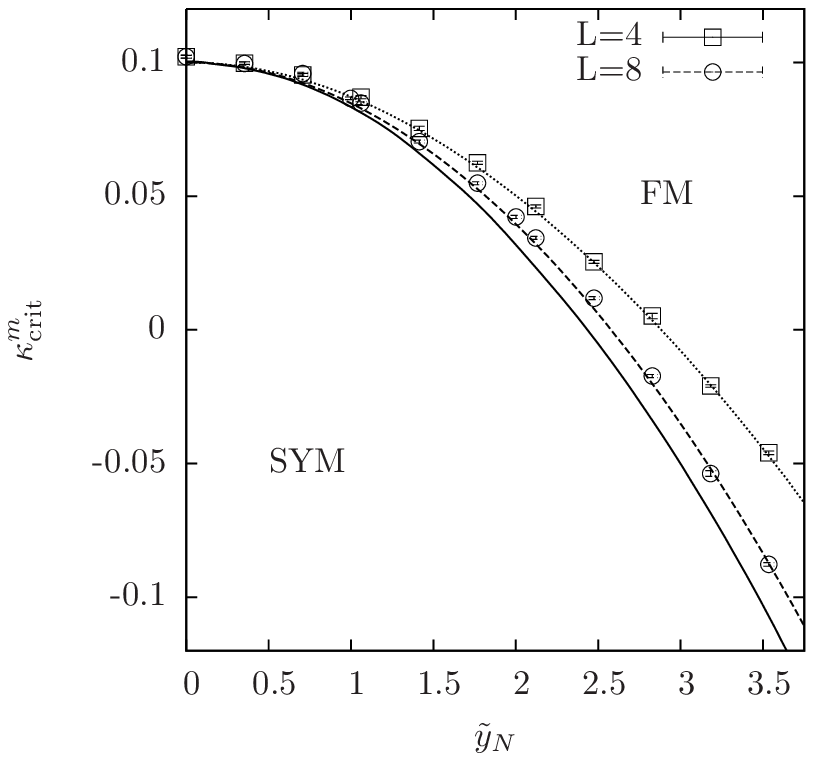}
\hs{5mm}&
\hs{5mm}
\includegraphics[width=0.32\textwidth]{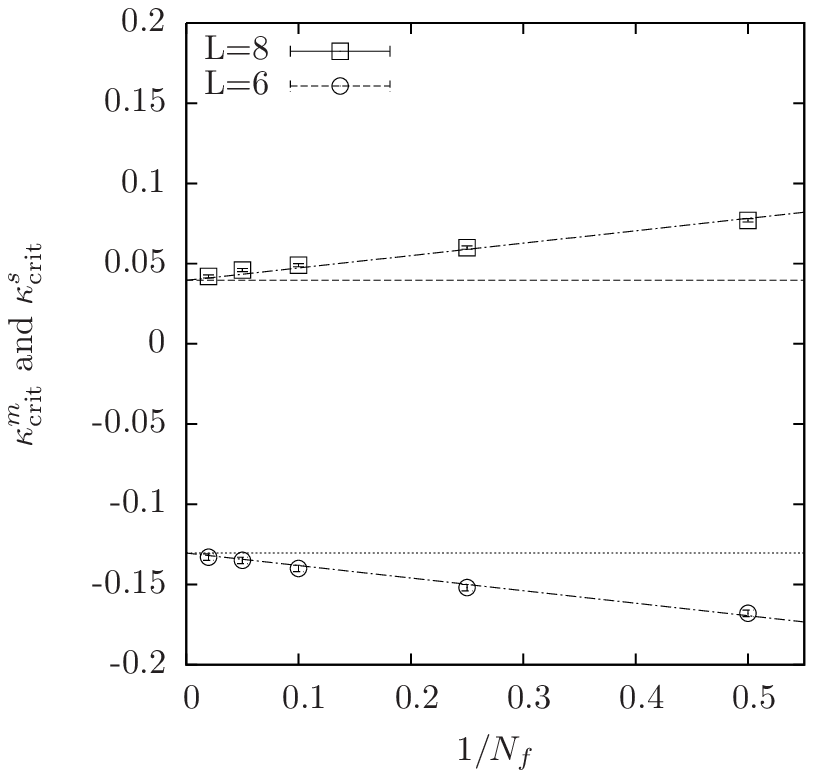}
\hs{5mm}\\
\hs{15mm}(a) & \hs{15mm}(b)  \\
\end{tabular}
\caption{(a) Some selected phase transition points between the ferromagnetic and 
the symmetric phase, as obtained at $N_f=50$ on a $4^4$-lattice (open squares) and on an 
$8^4$-lattice (open circles), are compared to the $L=4$ (dotted), 
$L=8$ (dashed), and $L=\infty$ (solid) phase transition lines determined analytically
in the large $N_f$-limit.
(b) $N_f$-dependence of $\kapCrit^m$, $\kapCrit^s$ at $\tilde y_N=2.0$ as
obtained on an $8^4$-lattice (open squares) and on a $6^4$-lattice (open circles).
The analytical, finite volume, large $N_f$ predictions for the SYM-FM (SYM-AFM)
phase transitions are represented by the dashed (dotted) lines. 
The dash-dotted lines are fits of the numerical data to linear functions as explained
in the main text.
In both plots $\tilde\lambda_N=0.1$ was chosen.}
\label{fig:Deviations}
\vs{-6mm}
\end{figure}

The finite size effects are illustrated in Fig.~\ref{fig:Deviations}a, showing
some phase transition points from the FM to the SYM phase as obtained 
from our numerical simulations on a $4^4$-lattice (open squares), and on an $8^4$-lattice 
(open circles) for the (very large) value
of fermion generations $N_f=50$, chosen to isolate the finite size effects from the
$1/N_f$ corrections. One clearly observes that the phase transition line is shifted 
towards smaller values of the hopping parameter when the lattice size is increased. 
The numerical results are compared to the $N_f=\infty$
phase transition lines obtained for
\bc
\setlength{\unitlength}{0.01mm}
\begin{figure}[t]
\begin{tabular}{cc}
\begin{picture}(6600,5500)
\put(600,500){\includegraphics[width=5cm]{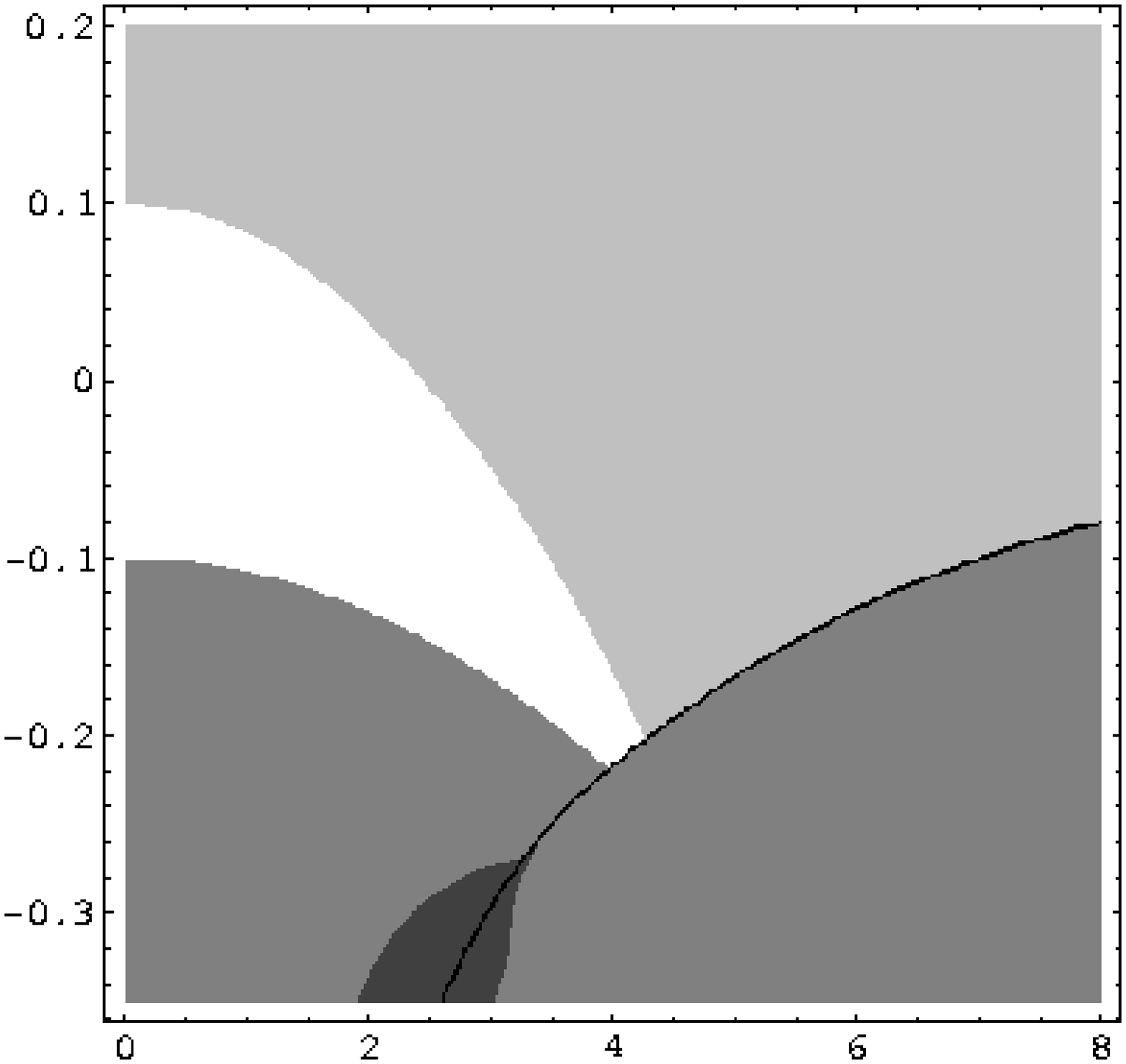}}
\put(25,3100){$\tilde \kappa_N$}
\put(3200,300){$\tilde y_N$}
\put(1300,3350){\textbf {SYM}}
\put(4500,3350){\textbf {FM}}
\put(1300,1500){\textbf {AFM}}
\put(4500,1500){\textbf {AFM}}
\put(2900,1000){$\leftarrow$\textbf {FI}}
\end{picture}
&
\begin{picture}(6600,5500)
\put(600,100){\includegraphics[width=5.9cm]{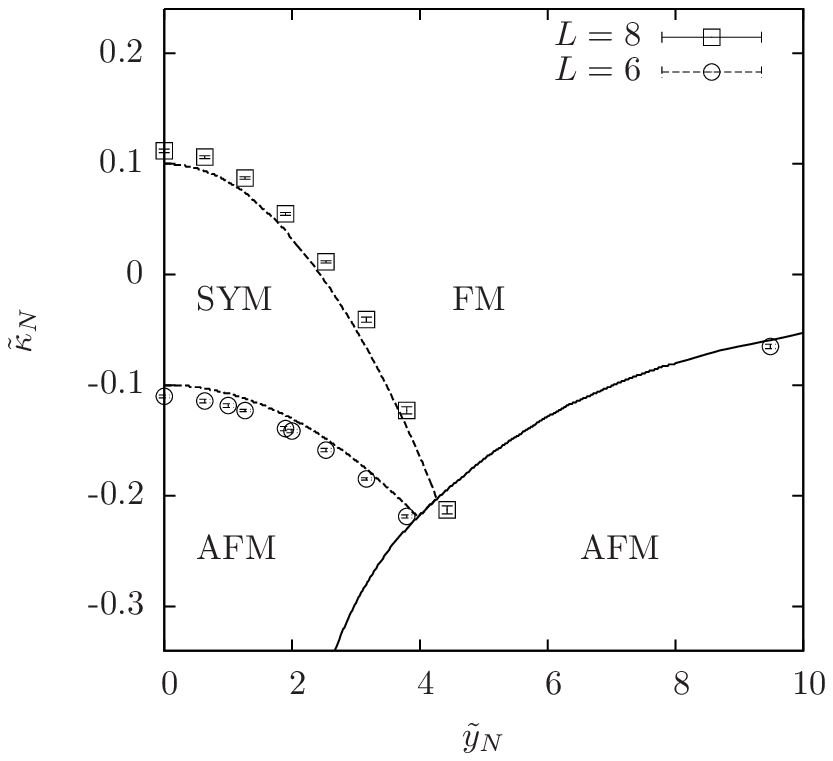}}
\end{picture}
\\
(a) & (b)
\end{tabular}
\caption{Phase diagrams with respect to the Yukawa coupling constant $\tilde y_N$ and the hopping 
parameter $\tilde \kappa_N$ for the constant quartic coupling $\tilde \lambda_N=0.1$.
The black solid line indicates a first order phase transition, while the remaining
transitions are of second order~\cite{Gerhold:2007gx}.
(a) Analytically obtained phase diagrams for $L=\infty$ and $N_f=\infty$.
(b) Comparison with numerically obtained phase transition points for $N_f=10$
and $L^4=8^4$ (open squares) and $L^4=6^4$ (open circles).
}
\vs{-10mm}
\label{fig:PhaseDiagrams1}
\end{figure}
\ec
\bc
\ec
$L=4$ (dotted line), $L=8$ (dashed line), and 
$L=\infty$ (solid line).
These analytically obtained lines perfectly describe the numerical results and one 
clearly observes the convergence of the numerical results to the analytically 
predicted $L=\infty$ line as the lattice size increases.

The $N_f$-dependence of the numerically obtained critical hopping parameters $\kapCrit^m$ and $\kapCrit^s$ is shown in Fig.~\ref{fig:Deviations}b for $\tilde y_N=2$ .
One clearly sees that for increasing $N_f$ the numerical results converge very well 
to the analytical finite volume predictions, as expected. It is interesting to note that
the leading term in the finite $N_f$ corrections, \ie the $1/N_f$ contribution, seems to 
be the only relevant correction here, even at the small value $N_f=2$, 
as can be seen in Fig.~\ref{fig:Deviations}b by fitting the deviations to the function 
$f_{m,s}(N_f)=A_{m,s}/N_f$ with $A_{m,s}$ being the only free parameter.
Furthermore, one observes that the critical hopping parameter $\kapCrit^m$ is shifted
towards larger values with decreasing $N_f$ while $\kapCrit^s$ is
shifted towards smaller values.

For an investigation of the model at large values of the Yukawa coupling constant
see Refs.~\cite{Gerhold:2007yb,Gerhold:2007gx}.

\section{Outlook towards Higgs mass bounds}
\label{sec:Outlook}

In contrast to the previous discussion, where we considered the model mostly in
the large $N_f$-limit, we now turn towards the physically interesting situation $N_f=1$.
In order to investigate the model also at odd values of $N_f$ we have implemented 
a PHMC-algorithm, which we will discuss in detail in an upcoming publication.

\begin{figure}[htb]
\begin{tabular}{cc}
\includegraphics[width=0.42\textwidth]{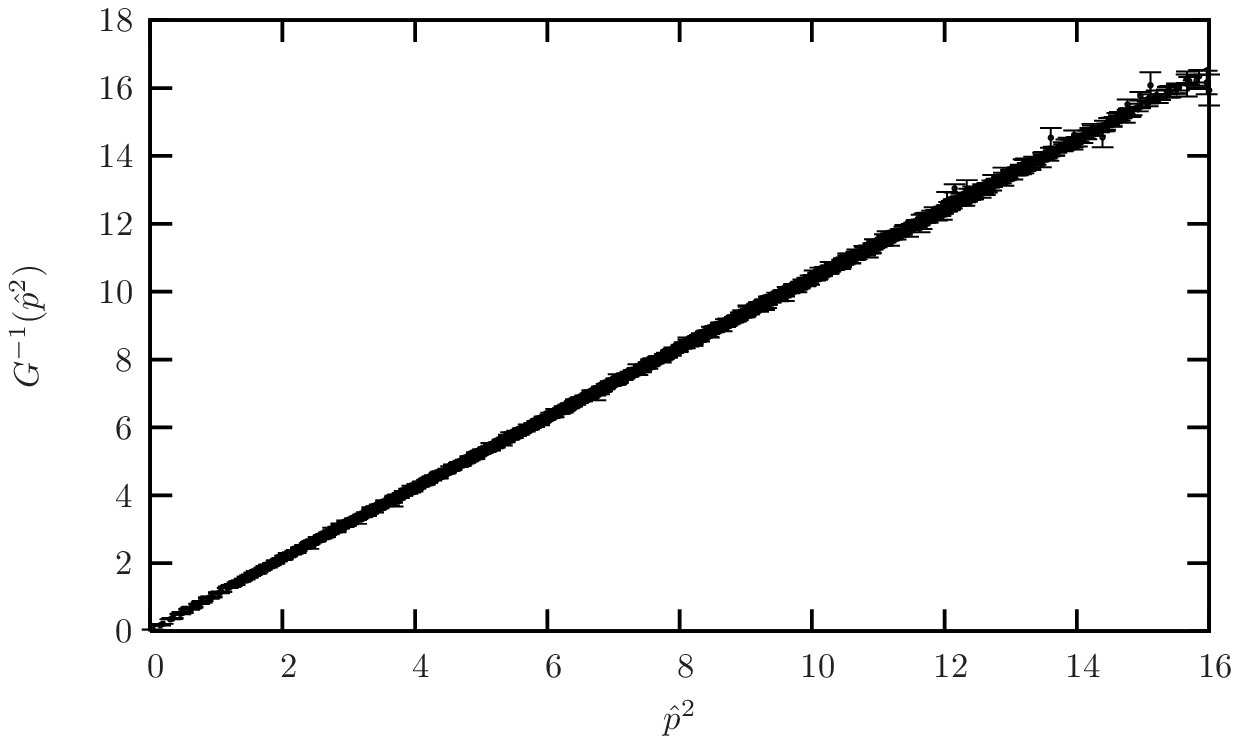}
&
\includegraphics[width=0.42\textwidth]{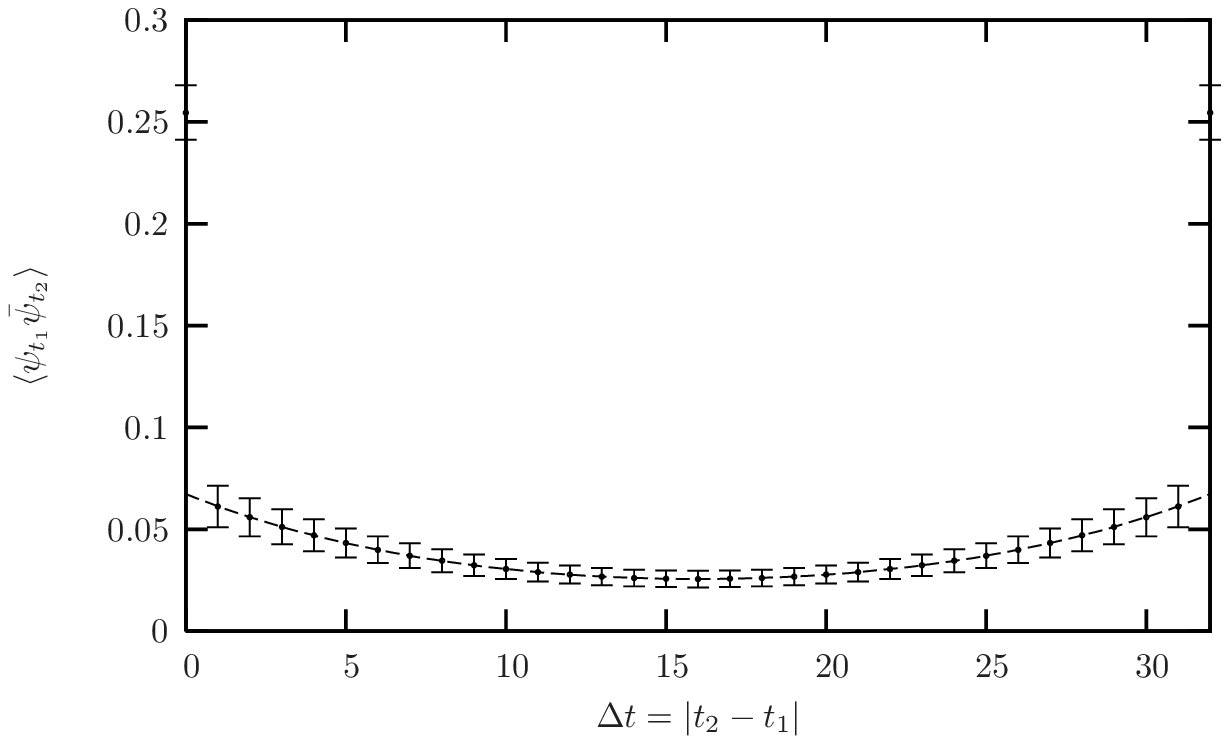}\\
(a) & (b)  \\
\end{tabular}
\caption{(a) Inverse Goldstone propagator $G^{-1}(\hat p^2)$ versus the squared lattice momentum $\hat p^2$
fitted to a linear function.
(b) Fermion time slice correlator $\langle \psi_{t_1}\bar\psi_{t_2} \rangle$ versus
distance in time direction $\Delta t = |t_2 - t_1|$ fitted to a $cosh$-function.}
\label{fig:Propagators}
\vs{-6mm}
\end{figure}

The main goal here is to compute the cutoff $\Lambda$-dependence of the
Higgs boson mass by fixing the top quark mass and the vacuum expectation 
value $v$ to their phenomenologically known values, \ie 
$m_\mathrm{top}=175\,$GeV and $v=246\,$GeV. From this dependence one can 
eventually determine an upper bound of the Higgs boson mass.
The $v$ measured on the lattice has to be renormalized
by the Goldstone renormalization factor $Z_G$ which can be obtained from the 
Goldstone-propagator $G(\hat p^2)$ according to
\beq
G^{-1}(\hat p^2) = \frac{\hat p^2}{Z_G}
\eeq
with $\hat p^2$ denoting the squared lattice momentum. For the chosen setting
($\kappa_N=0.240, y_N=0.711, \lambda_N=1.0$)
we obtain  $Z_G = 0.9662\pm 0.0001$ from the inverse Goldstone-propagator,
plotted in Fig.~\ref{fig:Propagators}a, and $\Lambda= (1684\pm 14)\,$GeV. In 
Fig.~\ref{fig:Propagators}b we show one selected component of the fermion
correlator $\langle \psi_{t_1}\bar\psi_{t_2} \rangle$ yielding the top mass
$m_{\mathrm{top}} = (170 \pm 6)\,$GeV in accordance with the phenomenological value.
\bc
\ec

\begin{figure}[htb]
\begin{tabular}{cc}
\includegraphics[width=0.42\textwidth]{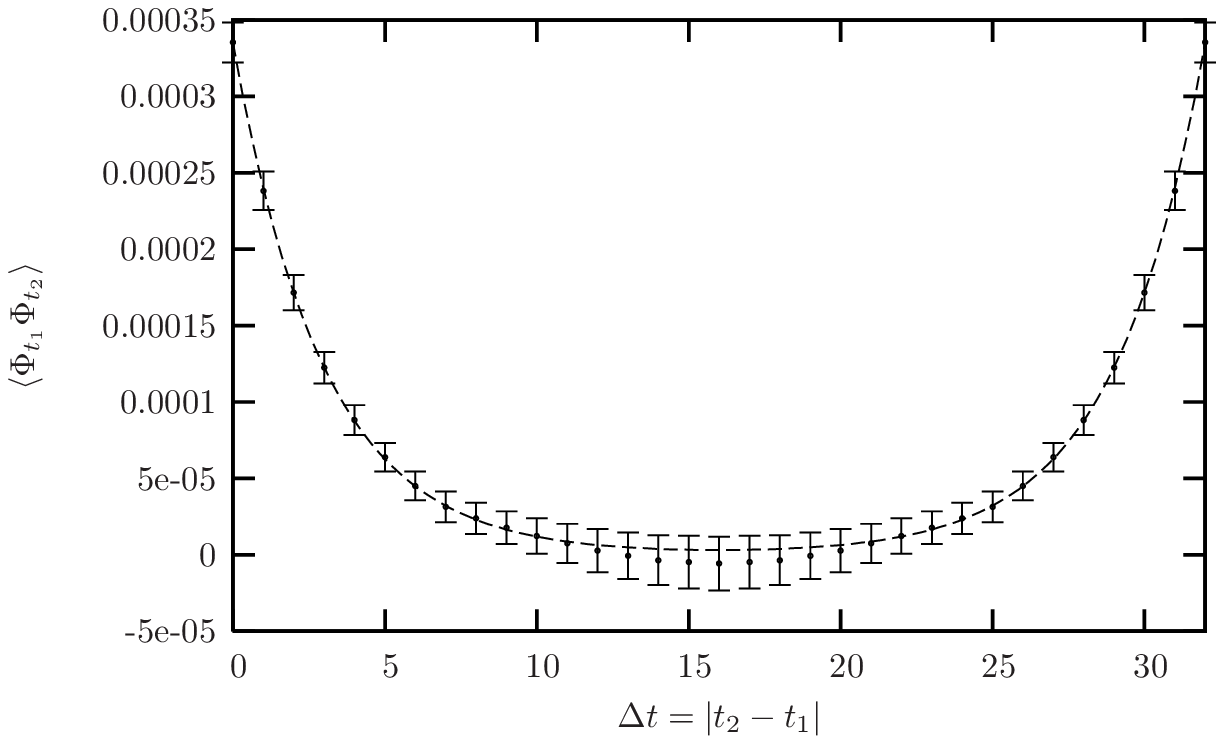}
&
\includegraphics[width=0.42\textwidth]{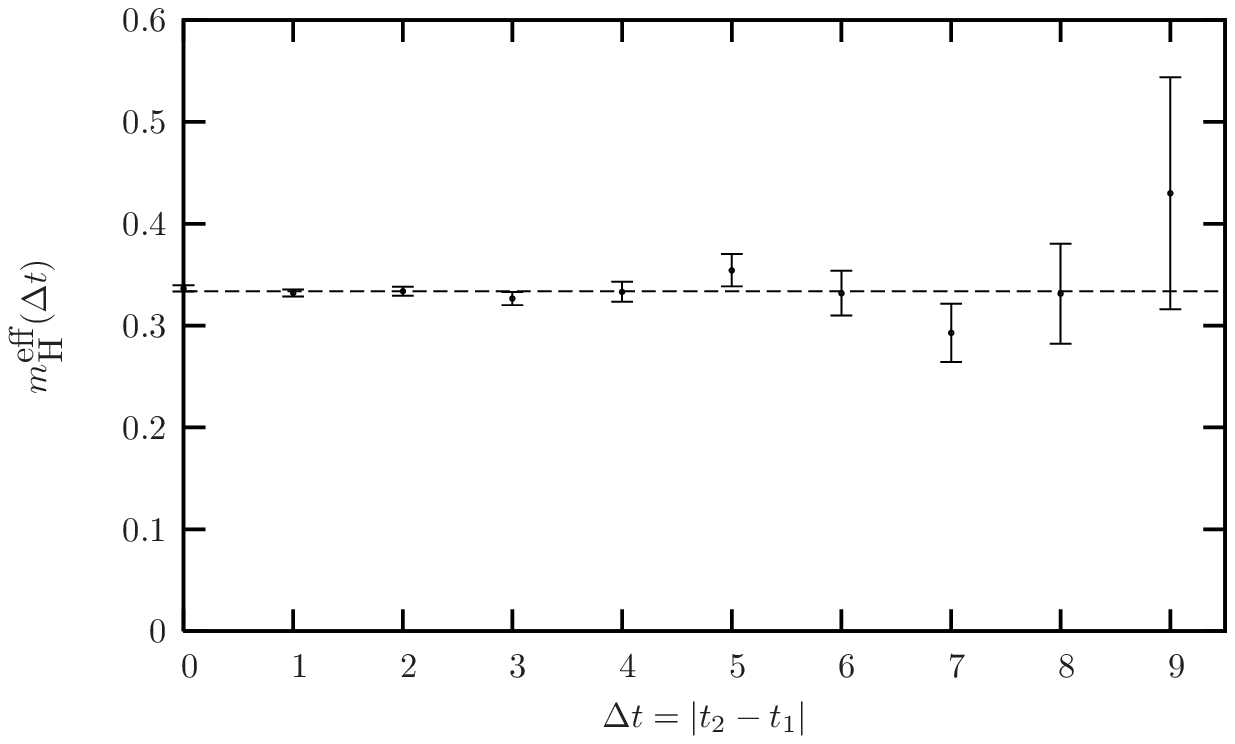}\\
(a) &
(b) \\
\end{tabular}
\caption{(a) Higgs time slice correlator $\langle \Phi_{t_1}\Phi_{t_2}\rangle$
versus $\Delta t$ fitted to a $cosh$-function.
(b) Effective masses $m_{\mathrm{H}}^{\mathrm{eff}}$ at $\Delta t$ fitted to plateau value $m_\mathrm{H}$.}
\label{fig:Masses}
\vs{-6mm}
\end{figure}

In the presented setup we chose the relatively large value of the quartic coupling
constant $\lambda_N=1$, aiming for an upper Higgs mass bound. In Fig.~\ref{fig:Masses}a 
we present the corresponding result for the Higgs correlator $\langle \Phi_{t_1}\Phi_{t_2} \rangle$
versus $\Delta t$. We determine the Higgs mass by calculating the effective mass
$m_{\mathrm{H}}^{\mathrm{eff}}$ at several values of $\Delta t$ and finding its plateau value
as shown in Fig.~\ref{fig:Masses}b. From this setup we find $m_\mathrm{H} = (565 \pm 15)\,$GeV.

However, we remark that here we give only a first and very preliminary result towards our
goal mentioned above. In particular, the value
$L\cdot m_\mathrm{top}=1.62$ is too small to determine the top quark mass reliably. Furthermore, the
statistics (2500 configurations for the Higgs analysis) is still to low to obtain sufficiently precise results
for the physical quantities of interest.

\bc
\ec

\section*{Acknowledgments}
We thank the "Deutsche Telekom Stiftung" for supporting this study by providing a Ph.D. scholarship for
P.G. We further acknowledge the support of the DFG through the DFG-project {\it Mu932/4-1}.
We are grateful to Joel Giedt, Julius Kuti, Michael M\"uller-Preussker, Erich Poppitz, and Christopher Schroeder
for enlightening discussions and comments. In particular we want to 
express our gratitude to Julius Kuti for inviting P.G. to his group at the University of
California, San Diego.

\nocite{*}
\bibliographystyle{unsrtOwnNoTitles}  
\bibliography{Proceedings}

\begin{thebibliography}{10}

\bibitem{Holland:2003jr}
K.~Holland and J.~Kuti.
\newblock Nucl. Phys. Proc. Suppl. 129, 765--767 (2004).

\bibitem{Holland:2004sd}
K.~Holland.
\newblock Nucl. Phys. Proc. Suppl. 140, 155--161 (2005).

\bibitem{Smit:1989tz}
J.~Smit.
\newblock Nucl. Phys. Proc. Suppl. 17, 3--16 (1990).

\bibitem{Shigemitsu:1991tc}
J.~Shigemitsu.
\newblock Nucl. Phys. Proc. Suppl. 20, 515--527 (1991).

\bibitem{Golterman:1990nx}
M.~F.~L. Golterman.
\newblock Nucl. Phys. Proc. Suppl. 20, 528--541 (1991).

\bibitem{book:Jersak}
A.~K. De and J.~Jers{\'a}k.
\newblock {HLRZ} {J\"u}lich, {HLRZ} 91-83, preprint edition (1991).

\bibitem{Luscher:1998pq}
M.~L{\"u}scher.
\newblock Phys. Lett. B428, 342--345 (1998).

\bibitem{Ginsparg:1981bj}
P.~H. Ginsparg and K.~G. {Wilson}.
\newblock Phys. Rev. D25, 2649 (1982).

\bibitem{Gerhold:2007yb}
P.~Gerhold and K.~Jansen.
\newblock JHEP 09, 041 (2007).

\bibitem{Gerhold:2007gx}
P.~Gerhold and K.~Jansen.
\newblock JHEP 10, 001 (2007).

\bibitem{Neuberger:1998wv}
H.~Neuberger.
\newblock Phys. Lett. B427, 353--355 (1998).

\end{thebibliography}

%\bibliographystyle{unsrtOWN}
%\bibliography{Proceedings}

%\begin{thebibliography}{99}
%  \bibitem{...} ....
%\end{thebibliography}

\end{document}